# Field emission properties of as-grown multiwalled carbon nanotube films

F. Giubileo[1,2,*], A. Di Bartolomeo[2,4], M. Sarno[3,4], C. Altavilla[3,4], S. Santandrea[2], P. Ciambelli[3,4] and A. M. Cucolo[1,2,3]

[1]*CNR-SPIN Salerno, via Ponte Don Melillo, 84084 Fisciano (SA), Italy*

[2]*Dipartimento di Fisica, Università di Salerno, via Ponte Don Melillo, 84084 Fisciano (SA), Italy*

[3]*Centro Interdipartimentale NanoMateS, Università di Salerno, via Ponte Don Melillo, 84084 Fisciano(SA), Italy*

[4]*Dipartimento di Ingegneria Chimica e Alimentare, Università di Salerno, via Ponte Don Melillo,84084 Fisciano(SA), Italy*

**Abstract:**

Multiwalled carbon nanotubes have been produced by ethylene catalytic chemical vapor deposition and used to fabricate thick and dense freestanding films ("buckypapers") by membrane filtering. Field emission properties of buckypapers have been locally studied by means of high vacuum atomic force microscopy with a standard metallic cantilever used as anode to collect electrons emitted from the sample. Buckypapers showed an interesting linear dependence in the Fowler-Nordheim plots demonstrating their suitability as emitters. By precisely tuning the tip-sample distance in the submicron region we found out that the field enhancement factor is not affected by distance variations up to 2μm. Finally, the study of current stability showed that the field emission current with intensity of about $3,3*10^{-5}$A remains remarkably stable (within 5% fluctuations) for several hours.

[*] Corresponding author: Tel: +39.089.969329, Fax: +39.089.969658. E-mail address: giubileo@sa.infn.it (Filippo Giubileo)

# 1. Introduction

Since Iijima's landmark paper on carbon nanotubes (CNTs) [1], great attention has been paid to their remarkable physical and mechanical properties such as high aspect ratio, chemical stability, high electrical conductivity and current carrying capability; their property of field induced electron emission (FE) has been higly emphasized for the relatively low threshold voltage, good emission stability and long emitter lifetime [2,3].

In the last decade, huge activity has been developed in the scientific community to investigate and apply the field emission properties of various carbon nanotube structures: CNT films differing for tube type, multiwalled or singlewalled, shape, dimension, density, substrate, etc. have been extensively studied for FE devices, like flat panel displays [2,4], cold cathodes [5], electron guns [6], X-ray sources [7-11], single multiwalled CNT (MWCNT) emitters [12-15], microwave power tube amplifiers and electron microscopy [16-18], etc.

Due to the inhomogeneous composition and morphology of CNT films, even on small areas, the study of the influence of fabrication and structural parameters on the FE properties represents a scientific challenge.

Various techniques have been used to fabricate CNT-based FE devices including direct synthesis of CNTs on device cathodes by chemical vapor deposition (CVD) [19]; realization of CNT-based paste where organic binder and vehicle are mixed with CNTs [20]; dielectrophoresis deposition [21, 22], dipcoating [23], and spraying [20]. All these techniques do not easily yield large size cathodes. The CNT mixture paste is currently the technique mostly used for field emitter fabrications, despite of its poor control on FE uniformity.

A CNT based structure called "buckypaper", which is a paper-like sheet of randomly oriented CNTs (where van der Waals interactions at the tube–tube intersections creates a freestanding film with properties similar to those of the constituting CNTs) and its composites have become a hot topic in CNT research and have been widely reported in recent years [25-36]. The intrinsic properties of buckypapers make them very useful for a broad range of applications: radio frequency filters [37], cold-field cathode emitters [38], etc. The flexibility and structural integrity of buckypaper has been also used for the production of artificial muscles [39] or of filtration devices exploiting the assembly of pores among the tubes [40].

Buckypaper contains several advantageous properties, such as extreme simplicity of fabrication and usability, mass production of large-scale samples, dense nanotube loading, and vast contacts of nanotubes/nanotube ropes; in this context buckypaper could also be one of the best candidate to realize large area cathodes with CNT emitters.

In this paper we report a detailed study of the field emission properties from as-grown MWCNT buckypaper. We first describe the fabrication process of the samples and their electrical characterization. Then, we study the field emission properties of buckypapers in special setup, where metallic cantilever in a scanning probe microscope is used as anode to collect electron current from small sample area. Performance of long duration current stability are also tested.

**2. CNT Buckypaper fabrication**

Multiwalled carbon nanotubes (MWCNTs) was synthesized by a catalytic chemical vapor deposition in an experimental plant equipped with on-line ABB analysers, that permit the monitoring of $C_2H_4$, $C_2H_2$, $CH_4$ and $H_2$ concentrations in the effluent stream on line during the reaction. Co, Fe catalysts (2.5 wt% of each metal) were prepared by dry impregnation with a cobalt acetate and iron acetate solution of gibbsite ($\gamma$-Al(OH)3). The catalyst was dried at 393 K for 720 min, and preheated before synthesis at 70 K/min up to 973 K under $N_2$ flow. For the CNT synthesis a mixture 15% v/v of ethylene in nitrogen was fed to a continuous flow microreactor, with a runtime of 30 min. Gas flow rates were controlled by calibrated Brooks mass flow controllers. Gas flow rate ad catalyst mass were 120 (stp)cm3/min and 400 mg respectively. The operative conditions were very effective ensuring to obtain selectively carbon nanotubes by converting the 98% of the ethylene fed to the reactor. Long bundles of carbon nanotubes were obtained (Figure 1a). MWCNT bundles produced (Figure 1b) have a length in the range 400-500 µm, internal and external diameter of tubes are 10-30 nm and 5-10 nm respectively.

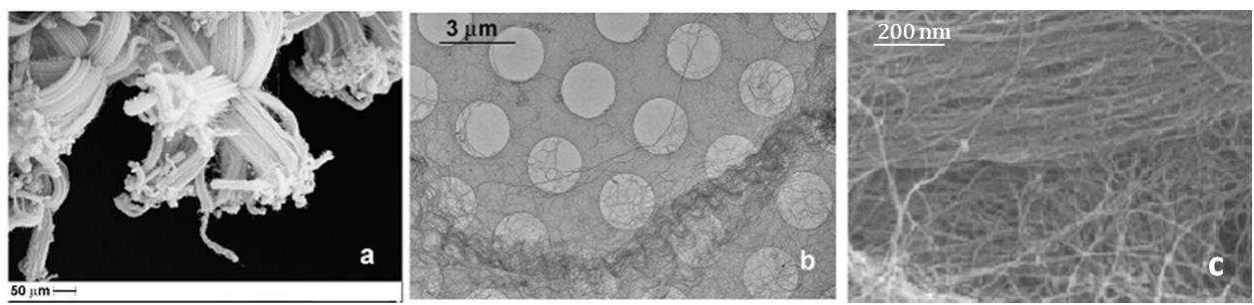

**Figure 1:** *(a) Scanning electron microscopy (SEM) image of MWCNTs bundles obtained with a LEO 1525 microscope.; (b) Transmission electron microscopy (TEM) image of MWCNT bundles obtained with a Jeol 1200 EX2 microscope. The preparation of samples for TEM observation involved sonication in ethanol for 2-5 minutes and deposition of the sample on a carbon grid.; (c) SEM image of a MWCNT buckypaper.*

Such MWCNTs, after a purification in HF to remove catalyst, were then used to fabricate CNT buckypapers [36]. Films of different thickness and densities were removed from the support, the thickness and density of CNTs in the films are easily controllable.

The so-obtained films can be folded and cut, being robust enough to let stable silver paint contacts to be formed and to withstand long thermal stresses. We point out also that, despite of its simplicity, this method gives freestanding buckypapers with a stable resistance without the application of a cross-linking agent to prompt a cross-linking reaction between carbon nanotubes [41]. Figure 1c shows SEM image of the buckypaper at different magnifications evidencing the formation of network of nanotubes bundles.

The fabrication process does not affect the characteristics of the MWCNTs as is demonstrated by Raman spectra measured at room temperature, with a microRaman spectrometer Renishaw inVia with 514 nm excitation wavelength. The Raman spectra for MWCNTs are dominated by two Raman lines at about 1590 cm$^{-1}$ (G-line) due to the in-plane vibration of the C-C bonds, and at about 1300 cm$^{-1}$ (D line) attributed to disorder induced by defects and curvature in the nanotube lattice. Together, these bands can be used to evaluate the extent of any carbon-containing defects. The D/G intensity ratios of the as-produced MWCNTs and that of the buckypaper are respectively 0.83 and 0.84, indicating that the treatments to produce buckypapers do not lead to an increase of the CNTs defects.

*2.1 Buckypaper electrical characterization*

A standard four-probe method was used to measure the low resistance of the MWCNT buckypapers, typically a few ohms, and to avoid the problem of the relatively high wire and contact resistance. The electric contacts were generally realized by silver paint. We measured several samples in order to get statistically significant information. For all samples we measured the temperature dependence of the resistance, R(T), over a wide temperature range (4.2K and 420K), and we also checked the current-voltage characteristics at different temperatures. The R(T) curves were measured by forcing a current through the outer probes and measuring the voltage across the inner ones. The forced current was kept as low as possible (normally below 0.1 mA) to prevent sample self-heating. In Figure 2 we report an example of the R(T) curve measured for one of our samples. We can observe a monotonous behavior where the buckypaper resistance increases for lowering temperature following an exponential dependence. Hovever, the low-voltage I-V curves are linear and this ohmic behavior is independent of the temperature as demonstrated by the three characteristics reported in figure 2, measured at T=4.2K, T=77K and T = 298 K respectively. We have already shown in a previous paper that a buckypaper can be used as small-size thermistor [36]

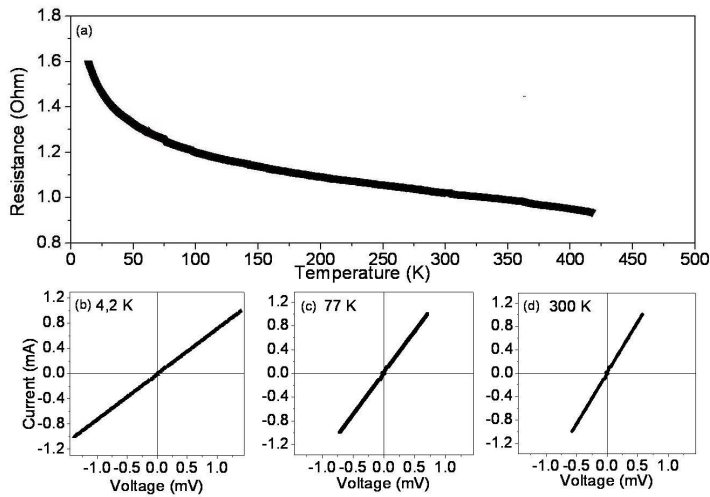

**Figure 2:** *(a) Resistance versus Temperature dependence of MWCNT buckypaper measured by 4-probes method in the temperature range between 4.2K and 420K. Current-Voltage characteristics are recorded at different temperatures for small voltages: (b) T=4.2K (c) T=77K (d) T=300K.*

## 3. Field Emission from MWCNT Buckypaper

To study field emission properties we arranged a special apparatus, by interconnecting a Semiconductor Parameter Analyser Keithley 4200-SC working as source-measurement unit (SMU) to room temperature high vacuum Atomic Force Microscopy system (schematic representation is drawn in Figure 3a). The buckypaper sample was placed on the sample holder of the Microscope while the metallic cantilever probe was used as second electrode. We used the SMU to apply a voltage between tip and sample in the range ±210V. The measurements of the field emission current versus the applied voltage were carried out under low pressure P≈$10^{-8}$ mBar. The high spatial resolution provided by the AFM piezoelectric controller allowed precise estimation of the tip-sample distance in the submicron region, with a resolution better than 1nm over a range of 2μm. The starting point to realize a field emission measurement was to approach the Pt–Ti coated cantilever to the buckypaper sample in the non-contact AFM mode by means of the automatic feedback system. In non-contact mode, the system vibrates the cantilever near its resonant frequency (about 300 kHz) with an amplitude of a few tens of angstroms and detects changes in the resonant frequency or vibration amplitude as the tip comes near the sample surface. The sensitivity of this detection scheme provides sub-nanometer vertical resolution in the movements. In this way, owing to the small force (1–10 nN) applied by the tip, it is possible to approach the tip to the sample surface gently enough to avoid morphological modifications . Once the automatic approach is completed, the probe can be retracted up to 2μm far from the buckypaper surface in order to start the FE characterization of the device. We notice that this procedure allows to estimate the tip-sample distance within 1-2 nanometers.

Correlation between the local topography and the field emission property was tried, but we didn't observe any clear dependence since we found similar FE results on several positions across the sample area.

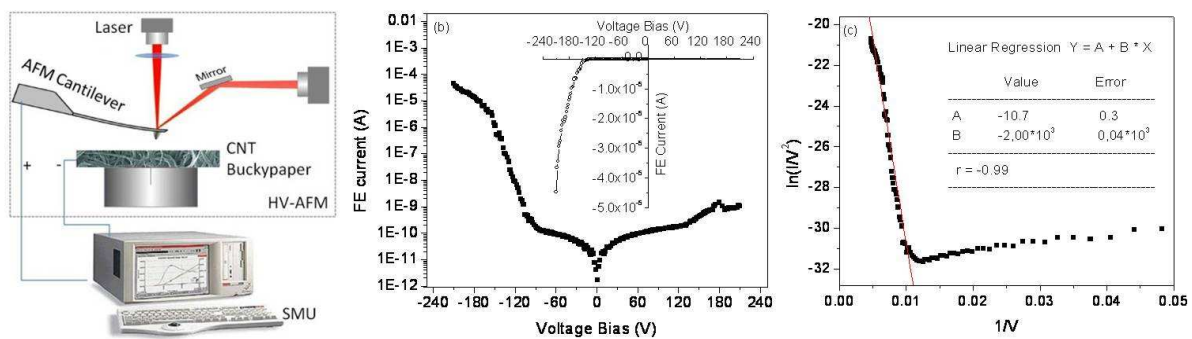

**Figure 3:** *(a) Schematic representation of the FE setup where SMU is interconnected to the AFM to apply voltage difference between tip and sample. (b) FE Curren-Voltage characteristic (logarithmic scale) measured at distance d=1μm (inset: linear scale); (c) FN plot for data reported in Figure 3b. Solid line represents linear fitting resulting in the parameters indicated on the graph.*

In Figure 3b we report a FE characteristic (I vs V) measured in the voltage range between -210V and +210V. The SMU system is capable to apply a voltage bias between tip and sample and simultaneously to measure the current with a sensitivity of the order of 0.1nA. The experimental data are shown in the logarithmic scale to evidence the sudden strong enhancement, of about 6 order of magnitudes, in the current intensity once field emission is triggered. We notice that the micrometric inter-electrode distance allows also to achieve high electric fields (up to 200 V/μm) with a modest voltage and attain current densities larger than $10^4$ A/cm$^2$.

To analyze field emission characteristics we refer to the usual Fowler–Nordheim (FN) theory[42], which relates the emission current density J to the local electric field ($E_S$) at the emitter surface and to the work function ($\varphi$) of the material, as expressed by the equation

$$J = \frac{I}{A} = a\left(E_S^2/\varphi\right)\exp\left[-b\varphi^{3/2}/E_S\right]$$

where $a$= 1.54·10$^{-6}$ AV$^{-2}$ eV and $b$=6.83·10$^7$ eV$^{-3/2}$Vcm$^{-1}$ are constant, I is the current and $A$ is the emitting surface area. The local electric field can be expressed in terms of the applied potential V and the cathode-anode distance $d$

$$E_S = \beta \cdot \left(\frac{V}{d}\right)$$

where β is the field enhancement factor due to emitter geometry. From the above expression it is possible to obtain a linear relation between ln(I/V$^2$) versus 1/V (FN plot) with slope given by

$$m = \left(b\varphi^{3/2}d\right)/\beta$$

This formula is properly true for standard parallel plate configuration. However, due to the special geometry of our setup, sensibly different from the standard parallel plate configuration, we need to take into account that the positively biased electrode is represented by the metallic probe of the atomic force microscope, precisely a Pt–Ti coated cantilever probe with curvature radius of about 50nm. Hence, it is necessary to introduce a further enhancement factor that depends on the tip shape through the curvature radius of its apex and a so-called tip correction factor $k_{eff}$. It has been shown [43] that in such a case the formula of the slope in the FN plot can be reasonably approximated by

$$m = \left(k_{eff} b \varphi^{3/2} d\right)/\beta$$

The use of such a sub-micrometric electrode as anode in the FE measurements, on the other had, gives the opportunity to access local FE properties of the CNT buckypaper, being the measured current resulting from contribution from sample areas as small as 1μm$^2$ as already discussed and demonstrated in [43,44].

The plot in figure 3c shows the FN plot of the I-V curve of figure 3b. The linearity of the FN plot reveals that the IV-curves are governed by a conventional FN tunneling. Thus, the β factor can be easily calculated from the slope $m$ of the FN plot resulting by the linear fitting of the experimental data (solid line in Figure 3c). By assuming the work function φ of the CNTs to be 5eV, the same of graphite or fullerenes [45], $k_{eff}$ =1.6 [43], d=1μm, we found out $\beta \approx 50$. Remarkably, this value is comparable to those measured on vertically-aligned carbon-nanotube films in similar conditions [43]. Published values range from few hundreds to several thousands, but at inter-electrode distances orders of magnitude higher than ours [46–49]. The onset voltage required to draw an emission current of 1 nA was found to be about 100 V (turn-on voltage). Similarly, the anode voltage required to draw a fixed emission current of 1 μA was found to be 150 V (threshold voltage). It is important to notice that the field emission current exceed 10$^{-5}$A for bias larger of about 170V. Considering an emitting area of the sample of about 1μm$^2$, it is easy to calculate that the current density flowing in the device is larger than 10$^4$A/cm$^2$. Such high current density, the high field enhancement factor, and the observation of no conditioning/aging effects on all tested samples, confirm that buckypapers are promising for field emission devices.

*3.1 Tip-sample distance and Current stability*

We further characterized the FE performances of buckypaper by studying the field enhancement factor dependence on the electrode separation and also the current stability over the time.
By tuning the inter-electrode distance d by means of the precise AFM piezoelectric system we have recorded the FE characteristics (I vs V) for different tip-sample distances (in the range up to 2 μm). In Figure 4a we report current-voltage curves in logarithmic scale. We can clearly observe, as expected, that by increasing the distance d, the turn on voltage increases, and the voltage sweep yields lower field emission currents.

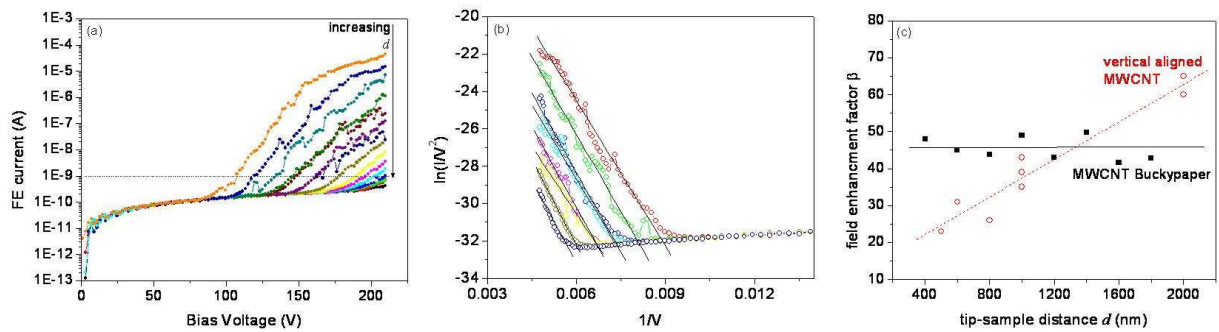

**Figure 4:** *(a) FE characteristics measured for different values of the tip-sample distance within the range between 0 and 2μm. (b) FN plots relative to data in figure 4a and corresponding linear fits (solid lines). (c) Field enhancement factor as function of tip-sample distance for buckypapers (full squares) is compared to the data previously obtained (empty circles) for vertically aligned MWCNTs [43].*

In Figure 4b we show the FN plots resulting from the experimental data of figure 5a. Solid lines represent the linear fitting for each curve, demonstrating once more that the field induced tunneling is responsible for the measured current. Moreover, from the slope of each plot we can determine the field enhancement factor, as we did for data of figure 3c, obtaining the dependence of β on the tip-sample distance, as shown in figure 4c. In such a figure the data (solid squares) are compared to the data (empty circles) previously measured in the case of FE from vertically aligned MWCNTs [42]. It is evident that in this distance range (400nm to 2000 nm) the behavior is significantly different, the enhancement factor being almost constant for the buckypaper against a linear increase for the vertically aligned MWCNTs. As expected, this confirms that aligned nanotubes have better performances as emitters than randomly arranged ones; however the preparation of vertically aligned CNT samples is more difficult and expensive.

One important issue in CNT-based emitters is about emission stability and lifetime. Applying a constant voltage, the emission current versus time I(t) has been measured in order to verify its stability and the aging of the cathode.

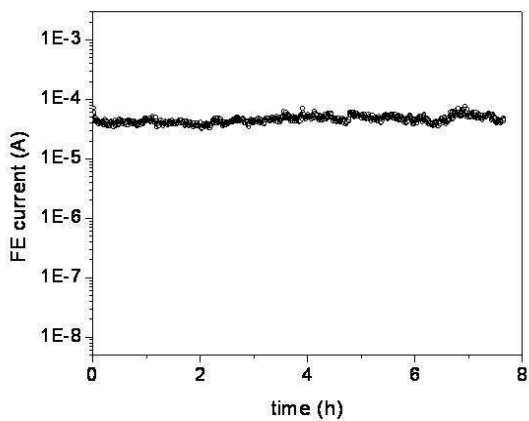

**Figure 5:** *Field Emission current stability of buckypaper measured by applying fixed bias of 150V.*

The long-term emission stability for MWCNT buckypaper has been measured at constant voltage of 150V. The current variation is reported in Figure 5 as a function of time. The buckypaper showed stable emission properties without degradation during operational time of about 8 hours. The emission current fluctuation is less than 5% around the average value $I=3,3*10^{-5}$A, confirming the high stability of the device.

## 4. Conclusion

In summary, MWCNT buckypaper have been fabricated and characterized. Special setup (based on AFM microscope) has been used in order to study field emission properties of as-grown buckypaper reducing the emitting area below $1\mu m^2$. The data are well described by a standard Fowler-Nordheim theory; the field enhancement factor $\beta$ resulted to be insensitive to the tip-sample distance within a range of $2\mu m$, differently from what was observed for vertically aligned MWCNT samples. The field emission currents have been found to be extremely stable over a period of about 8 hours with current intensities of about $3,3*10^{-5}$A. Our study shows that buckypaper is promising for FE application when high current but modest field enhancement is required.

**Figure Captions**

**Figure 1:** (a) Scanning electron microscopy (SEM) image of MWCNTs bundles obtained with a LEO 1525 microscope.; (b) Transmission electron microscopy (TEM) image of MWCNT bundles obtained with a Jeol 1200 EX2 microscope. The preparation of samples for TEM observation involved sonication in ethanol for 2-5 minutes and deposition of the sample on a carbon grid.; (c) SEM image of a MWCNT buckypaper.

**Figure 2:** (a) Resistance versus Temperature dependence of MWCNT buckypaper measured by 4-probes method in the temperature range between 4.2K and 420K. Current-Voltage characteristics are recorded at different temperatures for small voltages: (b) T=4.2K (c) T=77K (d) T=300K.

**Figure 3:** (a) Schematic representation of the FE setup where SMU is interconnected to the AFM to apply voltage difference between tip and sample. (b) FE Curren-Voltage characteristic (logarithmic scale) measured at distance d=1µm (inset: linear scale); (c) FN plot for data reported in Figure 3b. Solid line represents linear fitting resulting in the parameters indicated on the graph.

**Figure 4:** (a) FE characteristics measured for different values of the tip-sample distance within the range between 0 and 2µm. (b) FN plots relative to data in figure 4a and corresponding linear fits (solid lines). (c) Field enhancement factor as function of tip-sample distance for buckypapers (full squares) is compared to the data previously obtained (empty circles) for vertically aligned MWCNTs [43].

**Figure 5:** Field Emission current stability of buckypaper measured by applying fixed bias of 150V.